# Competitive interfacial charge transfer to graphene from the electrode contacts and surface adsorbates

Ryo Nouchi[1,a)] and Katsumi Tanigaki[2,3]

[1]Nanoscience and Nanotechnology Research Center, Osaka Prefecture University, Sakai 599-8570, Japan

[2]WPI-Advanced Institute for Materials Research, Tohoku University, Sendai 980-8577, Japan

[3]Department of Physics, Graduate School of Science, Tohoku University, Sendai 980-8578, Japan

[a)] E-mail: r-nouchi@21c.osakafu-u.ac.jp

ABSTRACT: Charge transfer (CT) at metal-graphene contacts induces a potential variation from the contact edges that extends to ~1 μm. Potential variations with a similar length should be observed around charge-transferring surface adsorbates. Thus, it is expected that a competition exists between these two CT sources when one source is within ~1 μm from the other. In this letter, weakly-coupled Ni contacts and 7,7,8,8-tetracyanoquinodimethan molecules are employed as the CT sources to investigate their possible competition. The CT from the molecules adsorbed only in the channel region change the charge density of the graphene in the under-contact regions. The extent of the CT effect in the under-contact region is as long as ~4 μm. The considerably long CT is ascribed to the high effective dielectric constant of the graphene under the contacts, resulting from a thin interfacial $NiO_x$ layer containing carbon impurities acquired from the graphene.

Graphene, a single layer of graphite, consists of carbon atoms arranged in a honeycomb lattice structure. All of the carbon atoms in graphene belong to its surface. Thus, the electronic properties of graphene are sensitive to surface/interfacial phenomena, making graphene promising for various sensing applications.[1] Interfaces with various materials are ubiquitously found in electronic/opt-electronic devices made of graphene. Among the various interfacial phenomena, charge transfer (CT) across the interfaces is of great importance to understand the electrical properties of graphene-based systems because it can greatly change the charge carrier concentration of graphene. One of the major sources of interfacial CT is metallic electrode contacts,[2,3] which exist in most electronic devices. The CT induced by the metal contacts occurs to align the Fermi levels of the two systems: graphene and metal. This type of CT affects the electrical properties of graphene field-effect transistors (FETs), evidenced by the manifestation of the electron–hole conduction asymmetry[4] and their enhancement by shortening the inter-electrode spacing.[5] The CT shifts the Fermi level of graphene under the metal contacts from the charge neutrality point, while the Fermi level of the region far from the contacts remains unchanged at the neutrality point. The electrostatic potential ($V(x)$) between these two regions varies by the functions of either $x^{-1/2}$ or $x^{-1}$ for undoped and doped graphene, respectively, where $x$ is the distance from the metal contact edge.[6] The length of the transition region has been reported to be as long as 1 μm.[7,8] This indicates that the effect of the CT extends to 1 μm from the point where CT takes place.

The source of the CT should not be limited to metal contacts. Among the various CT sources, other than metal contacts, surface adsorbates are widely used to dope charge carriers into graphene channels and are the most common type of analyte used when graphene FETs are operated as sensors. The surface adsorption of foreign atoms/molecules may occur at any position along the channel between the metal contacts. Thus, the metal contacts are within 1 μm from the adsorbates, where the CT effects from the two sources can compete with each other.

In this letter, we investigate the electrical characteristics of back-gated graphene FETs as a model system to study the possible competition between the two CT sources: metal contacts and surface adsorbates. The CT from surface adsorbates changes the carrier concentration of the graphene under the weakly-coupled metal contacts where the adsorbates are not in direct contact. From the measurements on the dependency on the length of the metal contacts (*d*), the extent of the CT from the surface adsorbates reached ~4 μm from the point where they were adsorbed. This is longer than the reported values for the CT effects from metal contacts and can be explained by the high effective dielectric constant of the graphene under the contacts.

Single layer graphene flakes were formed by mechanical exfoliation with adhesive tape onto Si substrates with a 300-nm thick thermal oxide layer grown on top. The Si substrates were highly doped and used as the back gate electrode. The number of layers of flakes was determined using optical contrast and Raman scattering spectroscopy. Electron-beam lithographic processes were used to pattern source and drain electrodes onto the graphene flakes. Ni was thermally evaporated under a high vacuum on the order of $10^{-4}$ Pa as the electrode metal, followed by a liftoff process. The thickness of the Ni electrodes was 20 nm. After the liftoff process, the substrates were rinsed in 2-propanol and acetone. The channel length of all the fabricated devices was 0.8 μm. 7,7,8,8-tetracyanoquinodimethan (TCNQ) molecules were used as the surface adsorbates. TCNQ is an electron acceptor[9] that dopes holes to graphene. Three mg of TCNQ (Aldrich, 98%, used as received) was dissolved in 150 ml of $CS_2$ (Wako, >99%, used as received). The TCNQ solution was dropped at room temperature onto the fabricated graphene FETs. The droplets almost entirely covered the surfaces of the substrates. The solution was then blown with air 10 s after dropping. Electrical characterization of the devices was conducted in air in the dark, at room temperature. During the measurements of transfer characteristics, the drain voltage ($V_D$) was set to 0.1 V.

Figure 1(a) shows the transfer characteristics of an as-fabricated Ni-contacted graphene FET. The contact length, the length of the source/drain electrodes along the channel length direction, of the device was 1.4 μm. The characteristics show a distorted curve, which is different from the commonly-observed V-shaped curve. The distorted transfer characteristics can be understood by the manifestation of charge-density depinning.[10,11] The resultant weak interactions led to double-dip-type transfer characteristics.[12] In the graphene FETs that displayed V-shaped transfer curves, the charge density of the graphene at the metal contacts could not be controlled with the gate voltage ($V_G$).[7] Conversely, for the electrodes made of metals that could be readily oxidized, the interfacial oxidation could de-pin the charge density at the metal contacts.[10,13] Note that annealing in a vacuum or in an inert atmosphere was not performed in the present study. Annealing has been widely employed to detach any unintentionally adsorbed impurities such as residues from the electron-beam resist, which is known to induce catalytic etching of the graphene under the metal contacts.[14] This type of etching might cause reconstruction of the interfacial chemistry. Direct metal-graphene coupling and pinning of the charge density should be restored at the metal contacts, as observed in the Ni-contacted graphene FETs annealed in an Ar atmosphere at 400°C for 12 hours.[10] Strong metal-graphene coupling pins the charge density of the graphene under the contacts and should protect the charge density from external disturbances such as gating.[7] In addition, first-principles studies at the level of the local density approximation[2,15] have suggested that strongly coupled contacts with metals such as Co, Ni, Pd, and Ti break the linear dispersion of the graphene through hybridization of the graphene $p_z$ states with the metal $d$ states. This type of metallization of the graphene chemisorbed on the metals is expected to hinder the modulation of the under-contact charge density by the CT effects from the nearby adsorbates. Thus, only weakly-coupled metal contacts with charge-density depinning will be treated in this study.

In the non-annealed Ni-contacted graphene FETs, interfacial oxidation started at the electrode edges and extended into the electrode center. The thickness of the interfacial oxide should be largest at the edge and smallest at the center, allowing charge carriers to be injected from near the center of the metal contact. This anomalous path that the carriers flow along was confirmed by the enhancement of the distortion in the transfer characteristics by increasing the length of the contact ($d$).[13] Thus, the distortion was attributed to the charge neutrality point of the graphene channel under the Ni contacts, whose length was approximately $d/2$. Figure 2 shows a potential profile along the path that the carriers flowed along. In this diagram, the positive (negative) potential indicates electron (hole) doping. The graphene channel under the Ni contacts was electron-doped because the distortion appeared in the negatively-gated region of the transfer curve (Fig. 1(a)).

After the adsorption of TCNQ, the graphene should be hole-doped. The metal electrodes act as a mask for the adsorption of TCNQ. Thus, the TCNQ molecules are only adsorbed along the channel. If the adsorbed TCNQ molecules dope holes only to the channel region (i.e., the adsorbed region), then the potential profile changes, as shown in the top panel in Fig. 2(a). The bottom panel in Fig. 2(a) illustrates the corresponding changes in the transfer characteristics, where the $V_G$ value corresponding to the charge neutrality point ($V_{CN}$) in the channel shifts toward higher $V_G$ values. In this case, the difference in the $V_{CN}$ values for the channel and under-contact regions ($\Delta V_{CN}$) becomes larger after the adsorption of TCNQ. Next, if the adsorbed TCNQ molecules also dope holes into the under-contact region, then the potential profile changes, as shown in the top panel in Fig. 2(b). The bottom panel in Fig. 2(b) illustrates the corresponding change in the transfer characteristics, where the $V_{CN}$ values of both the channel and under-contact regions shift toward higher $V_G$ values. This schematic is for an extreme case, where the amounts that the $V_{CN}$ values shifted are the same in these two regions. In this extreme case, the $\Delta V_{CN}$ value does not change by the adsorption of TCNQ.

Figure 1(b) shows the transfer characteristics measured after the adsorption of TCNQ. The measurement was performed with a graphene FET identical to the one measured in Fig. 1(a). The $\Delta V_{CN}$ value was 47 V before any TCNQ was adsorbed and it was almost unchanged after adsorption (only a small change of −4 V was observed). This result coincides well with the schematic in Fig. 2(b). Thus, for the Ni-contacted graphene FETs with $d$ = 1.4 μm, TCNQ molecules were only adsorbed into the channel region and holes were doped into the under-contact region. Charge carriers are believed to have been mainly injected from the center of the metal contact and the length of the under-contact channel can be approximated as $d/2$.[13] Therefore, this result indicates that the TCNQ molecules doped holes at a point at least 0.7 μm from the point where the TCNQ was adsorbed.

To check how the distant doping effect extended, the same measurements were performed with devices with longer contacts. Figure 3(a) shows the transfer characteristics measured before and after the adsorption of TCNQ. The horizontal axes are the gate voltage with respect to the $V_{CN}$ of the channel region. The devices with $d$ = 1.4 μm were identical to that shown in Fig. 1. The $V_{CN}$ shifted with the adsorption of TCNQ. The $\Delta V_{CN}$ values are plotted in Fig. 3(b) with their corresponding carrier concentrations against $d/2$. In contrast to the devices with $d$ = 1.4 μm, the devices with $d$ = 2.4 and 4.4 μm had an increase in $\Delta V_{CN}$ after the adsorption of TCNQ. These increases indicate that the doping strength of the TCNQ molecules adsorbed in the channel region was higher in the channel than in the under-contact regions. However, a shift in $V_{CN}$ in the under-contact regions was also observed with the longer-contact devices, although the amount that it shifted was less than that of the channel regions. Therefore, the longer-contact devices are believed to be between the two extreme cases shown in Fig. 2(b) and (c) (the under-contact region is either completely undoped or fully doped). Even for the devices with $d$ = 4.4 μm, the shift in $V_{CN}$ in the under-contact region still reached 40% of that of the channel region, where the $V_{CN}$ values for the channel and under-contact regions shifted from 5 to 87 V and −42 to −8 V, respectively. The corresponding ratio between the potential shifts reached 50%. Therefore, the CT effect from the surface adsorbates extended to more than 2.2 μm (half the contact length of 4.4 μm) from the point where they were absorbed. This is considerably longer than that observed at the metal contacts (~1 μm).[7,8]

Khomyakov *et al.* theoretically studied the potential variations near metal contact and derived a variational solution for undoped graphene:[6]

$$V(x) \approx \frac{V_B}{\left(\sqrt{x/l_S + \beta_2^2} + \beta_1 - \beta_2\right)^{1/2} \left(x/l_S + \beta_1^{-2}\right)^{1/4}}, \quad (1)$$

where $\beta_1 = 0.915$ and $\beta_2 = 0.128$. $V_B = V(0)$, which is the potential of the graphene at the contacts, $l_S$ is a scaling length, expressed as $1.3 \times 10^{-20} \times \kappa/|V_B|$ (nm), where $\kappa$ is the effective dielectric constant of graphene and is obtained by finding the average of the dielectric constants of in the environment, *i.e.*, the top layer ($\kappa_t$) and the bottom layer ($\kappa_b$), $\kappa = (\kappa_t + \kappa_b)/2$.[16] The bottom layer was $SiO_2$, and $\kappa_b = 3.9$. The top layer of graphene in the channel between the electrode contacts was air ($\kappa_t = 1$), giving $\kappa \sim 2.5$ for the graphene in the channel region. However, for the graphene under the contacts, the top layer should be an interfacial metal oxide ($NiO_x$ in this study), meaning that $\kappa_t$ will no longer be 1.

It was assumed that the functions obtained for the potential variation in Ref. 6 (incl. Eq. (1)) were also valid for the potential variation caused by an adsorbed molecular layer. In this study, the variation is discussed by taking the data with $d = 4.4$ μm as an example. The graphene region under the Ni contacts in the devices was electron-doped and $V_{CN} = -42$ V. Equation (1) is only valid for undoped graphene, thus it is not applicable to this case. The $V(x)$ for the doped graphene was relatively unchanged for the range where $|V(x)|$ was significantly higher than the doping level ($|\mu_F|$) and asymptotically behaved as:

$$V(x) \approx \frac{V_B |V_B| l_S}{2|\mu_F| x}, \quad (2)$$

for $|V(x)| \ll |\mu_F|$.[6] The doping level of $V_{CN} = -42$ V under the contacts corresponds to a Fermi level shift of 0.20 eV from the charge neutrality point. For $|V(x)|$, $V_B = V(0)$, which corresponds to a potential shift in the channel region, caused by the adsorption of TCNQ. This shift was 0.22 eV. In addition, the potential shift in the graphene under the contacts was 0.11 eV (50% of that of the channel). The observed shift under the contacts ($\delta V$) can be approximated by taking the average potential shift in the path that the current flows through under the contacts, *i.e.*, $0 \leq x \leq d/2$, and can be expressed as:

$$\delta V \approx \frac{\int_0^{d/2} V(x)dx}{d/2}. \tag{3}$$

For $d = 4.4$ μm, the $V(x)$ in the integral range was in between the undoped and doped limits of Eqs (1) and (2), respectively. Thus, an approximated function was obtained by fitting the curve with a double exponential function, as shown in Fig. 4, which is asymptotic to Eq. (1) for small $x$ values and to Eq. (2) for large $x$ values. Using the approximated function, $\delta V$ was calculated using Eq. (3), where $\kappa_t$ is a fitting parameter. $\kappa_t \sim 6700$ had the best fit for $\delta V = 0.11$ eV. The fitting results are shown in Fig. 4, indicating that the CT effect from the surface adsorbates extended to ~4 μm from the adsorption point.

The dielectric constant of NiO has been reported to be ca. 20 (Ref. 17) and 30 (Ref. 18) at $10^2$ Hz. These values are significantly lower than the obtained $\kappa_t$ of ca. 6700, but the obtained $\kappa_t$ should be characterized at zero frequency. The dielectric constants of most substances are frequency dependent and the presence of defective sites contribute to the increase in the dielectric constant at low frequencies. Here, an interfacial oxide is formed by the diffusion of oxygen from the edges of the contacts without any high-oxygen-pressure or high-temperature treatments. Thus, incomplete oxidation or non-stoichiometry might be expected in the interfacial oxide layer. In addition, the temperature increased when Ni was thermally deposited, causing the carbon atoms in the graphene to be dissolved into the Ni films.[14] Thermal treatments have recently been shown to corrode the graphene layer under electrodes made of Pd, Ti,[19] Ni and Co.[14] A similar catalytic etching of graphene by metallic particles at high temperatures was first reported with Ni[20] and occurs with various metals including Zn,[21] Ag,[22] Co,[23] Cr, Ti, Pd and Al.[24] Dissolved carbon atoms in the interfacial oxide act as impurities, increasing the dielectric constant. For NiO single crystals, the incorporation of Li with a 1.2-ppm weight ratio increased the dielectric constant at $10^2$ Hz from 30 to $10^4$.[18] Therefore, the high $\kappa_t$ value obtained (~

6700) can be explained by the defective sites originating from the incomplete oxidization and/or dissolved carbon atoms.

In summary, a possible competition between two CT sources, metal contacts and surface adsorbates, was investigated using back-gated graphene FETs with weakly-coupled source/drain contacts. CT from surface adsorbates in the channel region between the drain and source electrodes was confirmed to modulate the charge density of the graphene regions under the weakly-coupled source/drain contacts. The CT effect of the surface adsorbates extended to ~4 μm from the adsorption point, which is considerably longer than those of metal contacts (~1 μm).[7,8] This result can be explained by the high effective dielectric constant of the graphene under the contacts. The high dielectric constant was believed to originate from a defective/impurity-containing thin metal oxide layer formed at the metal-graphene interfaces. Weakly-coupled contacts with an interfacial oxide layer form with a wide range of metallic species[13] and the carbon atoms in the graphene may be dissolved into the as-deposited wetting metal films as impurities.[14] Thus, the far-reaching adsorbate-induced modulation of the charge density of the graphene under the contacts, which has been unveiled in the present study, should be widely observed in graphene FETs. This knowledge will be useful in developing a detailed understanding of the sensing response, where the adsorption of foreign atoms and molecules can be detected as changes in the FET characteristics.

This work was supported in part by the Special Coordination Funds for Promoting Science and Technology, and a Grant-in-Aid for Scientific Research on Innovative Areas (No. 26107531) from the Ministry of Education, Culture, Sports, Science and Technology of Japan; and by a research grant from the Kansai Research Foundation for technology promotion. The graphite crystal used was supplied by M. Murakami and M. Shiraishi.

**Figures**

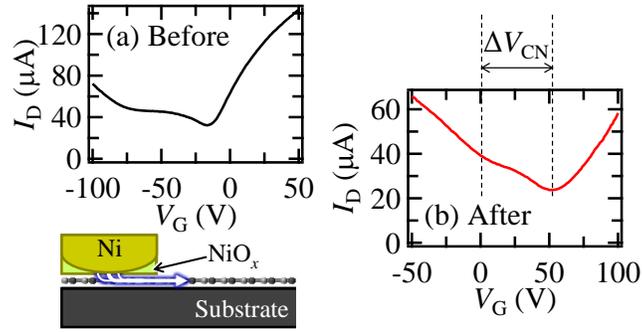

FIG. 1. Transfer (drain current ($I_D$) vs. gate voltage ($V_G$)) characteristics of an as-fabricated Ni-contacted graphene FET (a) before and (b) after the adsorption of TCNQ. The bottom panel of (a) is a schematic of the charge injection path. The contact length of the FET was 1.4 μm and the length of the current flow path under the contacts was approximated to be half the contact length, 0.7 μm. The difference in the gate voltages that corresponds to the charge neutrality point in the channel and contacted regions ($\Delta V_{CN}$) was almost unchanged after adsorption.

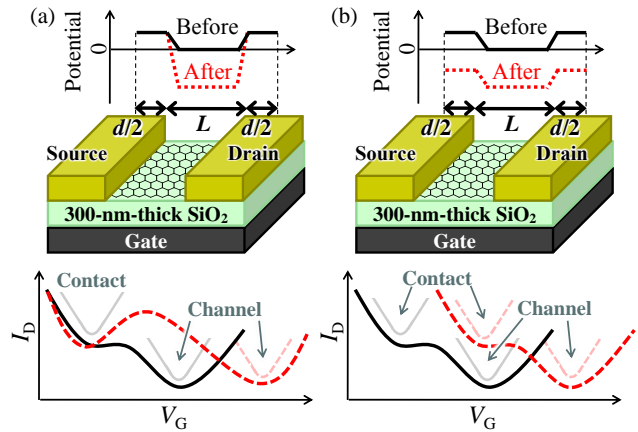

FIG. 2. Schematics of the potential profiles along the paths that the carriers flow along, and the corresponding transfer characteristics for the FETs before (solid lines) and after (dashed lines) the adsorption of TCNQ. Two extreme cases are considered: TCNQ molecules adsorbed into the channel (inter-electrode) region, (a) with only the channel doped and (b) with both the channel and the under-contact regions doped. After the adsorption of TCNQ, $\Delta V_{CN}$ increases in (a), but remains unchanged in (b). The positive (negative) potential indicates electron (hole) doping in these diagrams.

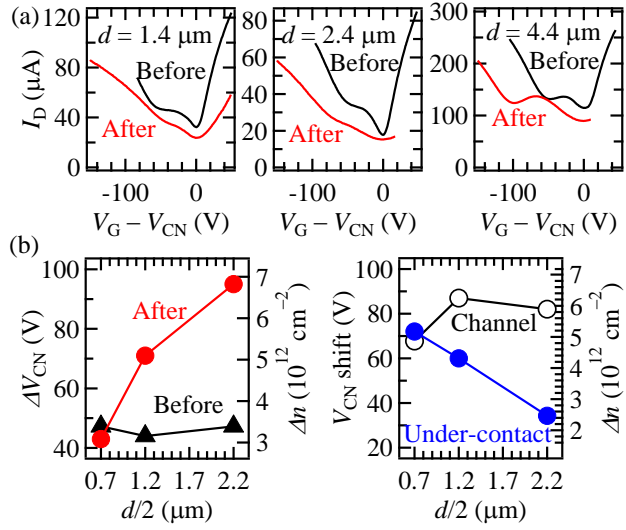

FIG. 3. (a) Contact-length dependence of the changes in the transfer characteristics induced by the adsorption of TCNQ. The horizontal axes are the gate voltage with respect to $V_{CN}$ of the channel region. The data for a contact length of $d = 1.4$ μm are identical to those shown in Fig. 1. (b) Extracted shifts in $V_{CN}$ before and after the adsorption of TCNQ and the $\Delta V_{CN}$ values with their corresponding carrier concentrations.

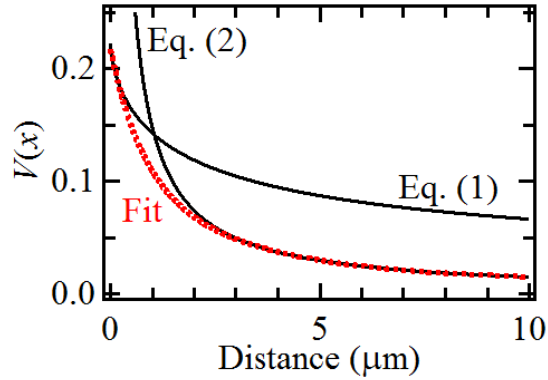

FIG. 4. Potential variations for a FET with $d = 4.4$ μm calculated using Eqs. (1) and (2) with $V_B = 0.22$ eV and $|\mu_F| = 0.20$ eV. The dashed line is a double exponential fit with $\kappa_t = 6700$, obtained under the condition where $\delta V = 0.11$ eV, calculated using Eq. (3).